\documentclass[aps,preprintnumbers,10pt,nofootinbib , twocolumn,amsmath,amssymb,prl,floatfix,superscriptaddress]{revtex4-2}
\usepackage[T1]{fontenc}\pdfoutput=1
\usepackage{times}
\usepackage{amssymb}
\usepackage{amsfonts}
\usepackage{amsmath}
\usepackage{microtype}
\usepackage{bm}
\usepackage{amsthm}
\usepackage{physics}
\usepackage{hyperref}
\hypersetup{colorlinks=true, citecolor=blue, urlcolor=blue}
\usepackage{dsfont}
\usepackage{physics}
\usepackage{cancel}
\usepackage{slashed}

\newcommand{\be}{\begin{equation}}
\newcommand{\ee}{\end{equation}}
\newcommand{\ba}{\begin{aligned}}
\newcommand{\ea}{\end{aligned}}

\def\mb{\mathbb}

\def\mc{\mathcal}
\def\bp{\begin{pmatrix}}
\def\ep{\end{pmatrix}}

\def\ptl{\partial}

\def\comma{\,,}
\def\period{\,.}

\def \be {\begin{equation}}
\def \ee {\end{equation}}
\def \bsp {\begin{split}}
\def \esp {\end{split}}
\def \bea {\begin{eqnarray}}
\def \eea {\end{eqnarray}}

\def\mc{\mathcal}
\def\ptl{\partial}
\def\mb{\mathbb}

\def \bl {\bar{\lambda}}

\begin{document}

\title{Understanding Fermionic Generalized Symmetries}

\author{Federico Ambrosino}
\email{federicoambrosino25@gmail.com}      
\affiliation{Deutsches Elektronen-Synchrotron DESY, Notkestr. 85, 22607 Hamburg, Germany}
\author{Ran Luo}
\email{ranluo@pku.edu.cn}
\affiliation{School of Physics,  Peking University, Beijing 100871, China}
\author{Yi-Nan Wang}
\email{ynwang@pku.edu.cn}
\affiliation{School of Physics,  Peking University, Beijing 100871, China}\affiliation{Center for High Energy Physics, Peking University,
 Beijing 100871, China}
 \author{Yi Zhang}
 \email{yi.cheung@pku.edu.cn}
 \affiliation{Center for High Energy Physics, Peking University,
 Beijing 100871, China}

\begin{abstract}
\noindent 
We explore new aspects of internal fermionic shifting symmetries, present in physical systems such as free Dirac spinors and $p$-form tensor-spinor fields. We propose a novel procedure to gauge these global symmetries, which also introduces a new St\"uckelberg mechanism to give a mass to free fermionic fields. Furthermore, we find new magnetic fermionic symmetries in these physical systems whose charged objects are disorder operators. For the case of a 4d Dirac spinor, we discuss a dual description, where the magnetic symmetry acts on the holonomies of a dual 2-form tensor-spinor. Further generalizations such as higher-group-like structures are also discussed.
\end{abstract}


\preprint{DESY-24-055}

\maketitle

\paragraph{\textbf{Introduction.}}

Symmetry is one of the most important guiding principles in the understanding of nature. In recent years, the study of generalized global symmetries has become a thriving field in theoretical physics, with applications in both high energy and condensed matter physics~\cite{Gaiotto:2014kfa,McGreevy:2022oyu,Schafer-Nameki:2023jdn,Brennan:2023mmt,Bhardwaj:2023kri,Luo:2023ive}. Comparing to the ``ordinary'' global symmetry group acting on local operators, one way to generalize is to consider \textit{higher-form symmetries} which act on extended operators by linked topological operators. Another generalization is to extend the symmetry group to more general algebraic structures such as higher-groups, non-invertible and higher-categorical symmetries.

Among the literature of generalized symmetries, the symmetry group/algebra themselves are almost always bosonic, such as finite groups, Lie groups and Lie algebras. In contrast, there has not been enough attention on the aspects of fermionic symmetries, despite of their appearance in many physical models. In this letter, we define a symmetry to be fermionic if the components of its symmetry parameter $\epsilon$ take value in the odd part of some Grassmann algebra\footnote{The symmetry group for fermionic symmetry can be taken as $\mb{R}^{0|s}$, which is the odd part of some supergroup. Note that the topology of such space is the discrete topology \cite{Rogers:1985yb,Rogers:2007zza,sachse2008categorical,DeWitt:2012mdz}, under which it is not sensible to define the notion of $U(1)$ or $\mb{Z}_N$ subgroups.}. For instance, global supersymmetry is a fermionic spacetime $0$-form symmetry, generated by the codimension-one topological operator $U_{\epsilon}(\mc{M}^{(d-1)})$ which is constructed with the supercurrent $J,\bar{J}$:
\be
U_{\epsilon}(\mc{M}^{(d-1)})=\exp\left(\int_{\mc{M}^{(d-1)}} i(\bar\epsilon \star J+\star\bar{J}\epsilon)\right)\period 
\ee
As an even simpler example, we can consider a massless Dirac spinor in a $d$-dimensional flat spacetime
\be
\label{diraclagrangian}
S = - \int \dd[d]{x} \,  \bar{\psi} \gamma^\mu \partial_\mu \psi\period 
\ee
This action is invariant under a $0$-form internal fermionic shifting symmetry $\psi\rightarrow\psi+\epsilon$, where $\epsilon$ is a constant spinor satisfying $\ptl_\mu\epsilon=0$. In \cite{Wang:2023iqt}, this was generalized to the notion of \textit{fermionic higher-form symmetries}, which can act on free fermionic $p$-form tensor-spinors, described by the action \eqref{tensorspinor}.

In this letter, we are going to investigate a number of fundamental physical questions for these internal fermionic symmetries. The first question is about the gauging thereof. It is well known that global fermionic symmetries can be sometimes gauged~\cite{Wang:2023iqt}, for instance, gauging global supersymmetry would lead to supergravity theories. Here, we propose a rather complete discussion of how to gauge fermionic $p$-form shifting symmetries of free tensor-spinor fields on flat-spacetime by \textit{minimal coupling} to $(p+1)$-tensor-spinor gauge field, with unconventional gauge transformation \eqref{p-form-modified-gt}, whose structure resembles the one of (A)dS supersymmetry, and indeed can be described as an In\"{o}n\"{u}-Wigner contraction of (A)dS superalgebra.  We present a physical model for our gauged system as a truncation of the  Volkov-Akulov (VA) model~\cite{Volkov:1972jx} with non-zero cosmological constant. 

The second question we address is the existence of magnetic symmetries for fermionic $p$-form tensor-spinors.  Bosonic $p$-form gauge fields in $d$ spacetime dimensions possess both an electric $p$-form symmetry and a dual magnetic $(d-p-2)$-form symmetry~\cite{Gaiotto:2014kfa}. The latter acts on disorder operators (such as the 't Hooft line operator when $p=1$), or on the holonomies of the electromagnetic (EM) dual $(d-p-2)$-form gauge field in the magnetic frame. This motivates us to revisit the long-standing question~\cite{Townsend:1979yv,Deser:1980fy} of dualization of fermionic fields. Here, we extend this to the fermionic case, illustrating that free $p$-form tensor-spinor fields also enjoy  \textit{magnetic} $(d-p-2)$-form fermionic symmetries whose charged object are disorder operators. For the specific case of $d=4, p=0$, we also present a \textit{dual magnetic} description, where the magnetic symmetry and its charged operators are represented as holonomies of dual \textit{magnetic} variables.

We conclude the letter by exploring fermionic algebraic structure beyond supergroup/superalgebra. We analyze a physical model for massive tensor-spinors in AdS spacetime, which has the structure of a gauged \textit{fermionic $2$-group symmetry}, for which we find a gauge-invariant \textit{fake curvature} in the system.  Finally, in Appendix B we comment on how to extend the discussion of this letter to curved spacetimes with covariantly constant spinors.

\paragraph{\textbf{Gauging shifting symmetry of Dirac spinors.}}
Consider a generic massless Dirac fermion $\psi$ in $d$ dimensions described by the standard action \eqref{diraclagrangian}, which is invariant under the global shifting symmetry:
\be \label{eq:Diracshiftvariation}
    \delta \psi = g\, \epsilon\comma 
\ee
by a constant spinor parameter $\epsilon$ with mass dimension $\frac{d-3}{2}$. $g$ is a constant with the dimension of mass, whose meaning will be clear in the following.
This symmetry is analogous to the shifting symmetry $\phi \to \phi + \theta$ of a free boson $\phi$. However, unlike the former case, the gauging of this fermionic $0$-form symmetry is non-standard due to the first order derivative nature of the Lagrangian \eqref{diraclagrangian}. 
Indeed, in order to promote the global shifting \eqref{eq:Diracshiftvariation} to a local symmetry,  one would be tempted to introduce a background vector-spinor gauge field $\lambda_\mu$, with prescribed gauge symmetry $\delta \lambda_\mu = \partial_\mu \epsilon$, and minimally couple by upgrading  partial derivatives $\partial_\mu \psi$ to their covariantized version $D_\mu \psi:= \partial_\mu \psi - g \lambda_\mu$ (and its left-acting version $\bar{\psi}\overleftarrow{D}_\mu:= - \partial_\mu \bar{\psi}  + g \bar{\lambda}_\mu)$. Yet, this would not lead to a gauge invariant action:
\be
 \delta \left( - \frac12 [\bar{\psi} \gamma^\mu D_\mu \psi + \bar{\psi}\overleftarrow{D}_\mu \gamma^\mu \psi] \right) = - \frac{1}{2} g  [\bar{\epsilon}\slashed{D}\psi + \bar{\psi} \overleftarrow{\slashed{D}} \epsilon] .
\ee
In order to construct a gauge-invariant system where the fermionic shift \eqref{eq:Diracshiftvariation} is gauged, we propose\footnote{An analogous modification was also proposed in \cite{Love:2003gm}, where the author considered in $d=4$ spacetime dimensions the gauging of the shifting symmetry of 2-components spinors, yet failing to construct a coupled system with dynamical gauge fields, invariant under simultaneous variations \eqref{eq:modifiedpsimuvar}.} the following modification of the gauge transformation of the background Rarita-Schwinger field as follows:
\be\ba
    \label{eq:modifiedpsimuvar}
    \delta \lambda_\mu = \partial_\mu{\epsilon} + \frac{g}{d} \gamma_\mu \epsilon &\comma \quad  \delta \bar{\lambda}_\mu = \partial_\mu{\bar{\epsilon}} - \frac{g}{d} \bar{\epsilon}\gamma_\mu \comma \quad \\
    \delta \psi = \alpha g \, \epsilon &\comma \quad \delta \overline{\psi} = \alpha g \, \overline{\epsilon}\comma
\ea\ee 
where $\alpha = \frac{\sqrt{(d-1)(d-2)}}{d}$. 
The unique action invariant under the gauge symmetry \eqref{eq:modifiedpsimuvar}, and that minimally couples the conserved 1-form currents:
\be
\mathcal{J}_\mu = \alpha g \gamma_\mu \psi \comma \qquad \bar{\mathcal{J}}_\mu = -  \alpha g  \bar{\psi} \gamma_\mu \comma 
\ee
to a dynamical Rarita-Schwinger gauge field, is:
\be
\begin{aligned} 
\label{0-form-gauged}
     S_{\rm\scriptstyle gauged} = \int\dd[d]{x}  \Big[ &- \bar{\psi} \gamma^\mu \partial_\mu \psi - g \bar{\psi} \psi -  \bar{\lambda}_\mu \gamma^{\mu\rho\nu} \partial_\rho \lambda_\nu \\
     &+ \frac{d-2}{d} g \bar{\lambda}_\mu \gamma^{\mu\nu} \lambda_\nu -  \bar{\lambda}_\mu {\mathcal{J}}^\mu  - \bar{\mathcal{J}}^\mu \lambda_\mu  \Big]  \period
\end{aligned}
\ee
This action necessarily contains mass terms for both the gauge field and Dirac field. Indeed, this gauged system can be interpreted as a fermionic equivalent to the \textit{St\"uckelberg mechanism} \cite{Stueckelberg:1938hvi,Stueckelberg:1938zz} describing a massive Rarita-Schwinger field with gauge symmetry \eqref{eq:modifiedpsimuvar}, where the Dirac spinor realizes the shifting symmetry \eqref{eq:Diracshiftvariation}. Upon fixing unitary gauge for the gauge field $\epsilon=-\psi/(\alpha g)$ in (\ref{eq:modifiedpsimuvar}), $\psi$ is set to zero and the action reduces to the action for an ordinary massive Rarita-Schwinger field\footnote{ Alternatively, fixing gamma-traceless gauge $\gamma^\mu \lambda_\mu = 0$ effectively decouples the two fields, and leads to a free massive Dirac and a free massive Rarita-Schwinger with an off-shell constraint.}. In this case, 
the degrees of freedom of $\psi$ are absorbed as longitudinal modes of the massive $\lambda_{\mu}$, and the fermionic gauge symmetry of $\lambda_{\mu}$ is broken by its mass term. This should be thought as the fermionic analogous of recovering the Proca equation by gauge fixing the St\"uckelberg action.

Finally, note that the coupling strength $g$ smoothly regulates a decoupling limit  $g\to 0$, where the theory reduces to a free massless Dirac spinor and a massless Rarita-Schwinger field with ordinary gauge symmetry $\delta\lambda_\mu = \partial_\mu \epsilon$.

\paragraph{\textbf{Gauging shifting symmetries of tensor-spinors.}}
The construction proposed above admits a direct generalization to the case of free fermionic $p$-form fields $\psi_{(p)}$.  It is convenient to express all the Lagrangians in a coordinate-free notation using the conventions in Appendix A for the $p$-forms spinors. 

A free action describing free antisymmetric tensor-spinor field of rank $p$ in $d$ spacetime dimensions ($d \ge 2 p + 1$) is a direct generalization of the Rarita-Schwinger action\footnote{In even dimensions, one has the freedom to add terms containing the chirality matrix $\gamma_{d+1}$. E.g.\ terms of the form $\bar{\psi}_{(p)}\wedge \gamma_{(d-2p-1)}\gamma_{d+1}\wedge d\psi_{(q)}$ still exhibit the shifting symmetry in \eqref{eq:p-formshifting}, (cfr. (3.5) of \cite{Wang:2023iqt}).}~\cite{Buchbinder:2009pa,Zinoviev:2009wh,Campoleoni:2009gs,Lekeu:2021oti}:
\be \label{tensorspinor}
S_{\rm \scriptstyle free}\left[\psi_{(p)}\right] = - \int_{\mathcal{M}^{(d)}}\bar{\psi}_{(p)} \wedge \gamma_{(d-2p-1)} \wedge \dd{ \psi_{(p)}} \period
\ee
This theory exhibits a global $p$-form shifting symmetry:
\be\label{eq:p-formshifting}
\delta \psi_{(p)} = \alpha g\, \epsilon_{(p)}\comma \qquad \dd{\epsilon_{(p)}} = 0\comma
\ee
generated by  the conserved $(p+1)$-form currents $\mathcal{J}_{(p+1)}$ and $\bar{\mathcal{J}}_{(p+1)}$ given by:
\begin{equation}
\begin{gathered}\label{currents}
    \star \mathcal{J}_{(p+1)} = \alpha g \, \gamma_{(d-2p-1)}\wedge\psi_{(p)}\comma\\
        \star \bar{\mathcal{J}}_{(p+1)} = -\alpha g \,  \bar{\psi}_{(p)}\wedge \gamma_{(d-2p-1)}\comma
\end{gathered}
\end{equation}
where $\alpha = \frac{\sqrt{(d-2p-1)(d-2p-2)}}{d}$. 
As a direct generalization of the ($0$-form) Dirac spinor story, to promote this global symmetry to a local one, we introduce a rank-$(p+1)$ antisymmetric tensor-spinor gauge field $\lambda_{(p+1)}$, that we must provide with the following modified gauge variation:
\be
\label{p-form-modified-gt}
\delta\lambda_{(p+1)} = \dd{\epsilon_{(p)}} + \frac{g}{d} \gamma_{(1)} \wedge \epsilon_{(p)}\period
\ee
Then, the unique minimally coupled gauged action is:
\begin{widetext}
\be
\begin{aligned} 
\label{p-form-gauged}
     S_{\rm\scriptstyle gauged} = S_{\rm \scriptstyle free}\left[\lambda_{(p+1)}\right] 
    &- (-1)^{p+1}\frac{g(d-2p-2)}{d} S_{\rm \scriptstyle mass}\left[\lambda_{(p+1)}\right] + S_{\rm \scriptstyle free}\left[\psi_{(p)}\right]  - (-1)^{p} \frac{g(d-2p)}{d}S_{\rm \scriptstyle mass}\left[\psi_{(p)}\right]   \\
     &-\int_{\mathcal{M}^{(d)}}  \Big[  \bar{\lambda}_{(p+1)} \wedge \star \mathcal{J}_{(p+1)}  + \star \bar{\mathcal{J}}_{(p+1)} \wedge \lambda_{(p+1)}  \Big]  \comma
\end{aligned}
\ee
\end{widetext}
where\footnote{For a version of this action in components, the reader can refer to \eqref{gaugedcationcurved} up to the replacement $\partial_\mu \leftrightarrow \nabla_\mu$.} the mass term for a $p$-form tensor-spinor is:
\be 
S_{\rm \scriptstyle mass} [\psi_{(p)}] = -\int_{\mathcal{M}^{(d)}} \bar{    \psi}_{(p)} \wedge \gamma_{(d-2p)} \wedge \psi_{(p)} \period
 \ee
It is clear that upon $p=0$, this action reduces to \eqref{0-form-gauged}, and all the considerations made with regards to the $0$-form gauging continue to hold here.

\paragraph{\textbf{Gauged theory from non-linearly realized SUSY.}}
\label{modifiedvariation}
The fermionic generators $S_\alpha$ of the global abelian shifting symmetry of the Dirac spinor satisfy the algebra (also known sometimes as \textit{internal supersymmetry}, cfr.\ e.g.\ \cite{Roest:2017uga}): 
\begin{equation} \label{eq:0formalgebra}
    \{S_\alpha, S_\beta\} = 0\comma \quad [M_{\mu\nu}, S_\alpha] = \frac{1}{2} (\gamma_{\mu\nu})_\alpha{}^{\beta} S_\beta \comma
\end{equation}
where $M_{\mu\nu}$ are the Lorentz generators and we omitted the non-vanishing commutation relations among them. 
This algebra can be obtained via In\"{o}n\"{u}-Wigner  (IW) contraction of the $\mathcal{N}=1$ super-Poincaré algebra \cite{Ivanov:1982bpa, Lopuszanski:1985ve, Roest:2017uga}, where the fermionic generators  $Q_\alpha$ are not internal symmetry generators but spacetime ones. This contraction amounts to introducing a continuous parameter $\omega$ such that $Q_\alpha= \omega S_\alpha$ and to take the limit $\omega \to \infty$, and reduces the generators of the spacetime SUSY in the $\mathcal{N}=1$ Poincaré algebra to interal fermionic generators satisfying \eqref{eq:0formalgebra}. 

The IW contraction we described above, can be realized by embedding the free Dirac field as \textit{small field  limit} (SFL) \cite{Roest:2017uga} of the Goldstone fermion in the Volkov-Akulov (VA) model \cite{Volkov:1972jx}.
The Lagrangian for the VA Goldstino $\lambda$ is expressed using the field $E^\mu_\nu = \delta^\mu_\nu + a^2 \bar{\lambda} \gamma^\mu \partial_\nu \lambda$ as:
\be
\mathcal{L}_{\lambda} = - \frac{1}{a^2} \text{det}(E) = - \frac{1}{a^2} - \bar{\lambda} \slashed \partial \lambda + \mathcal{O}(\lambda^3) \comma
\ee
where $a$ is the order parameter regulating the spontaneous global supersymmetry breaking. In the VA model, supersymmetry is only non-linearly realized on the Goldstino, that transforms as: $ \delta \lambda = \tfrac{1}{a} \epsilon + a (\bar{\epsilon}\gamma^\mu \lambda - \bar{\lambda}\gamma^\mu \epsilon) \partial_\mu \lambda$. In the SFL, once the Goldstino is appropriately scaled $\lambda \rightarrow \lambda/\omega$, the former symmetry transformation reduces to: 
$\delta \lambda =  \tfrac{1}{a} \epsilon$
\cite{Ivanov:1982bpa,Lopuszanski:1985ve,Roest:2017uga}. This, as expected, matches the global symmetry algebra \eqref{eq:0formalgebra} of the Dirac field shifting symmetry, and, in particular with \eqref{eq:modifiedpsimuvar} upon identifying the SUSY breaking scale with $a = 1/(\alpha g)$. As a matter of fact, one can  also check at the level of the Lagrangian description, that the VA model truncates, after IW contraction, to a free Dirac fermion in the presence of a constant $- 1/a^2$ term in the Lagrangian. 

Interestingly, the VA model also offers a concrete realization of the gauged system we propose \eqref{0-form-gauged}. 
Indeed, it is well established how to gauge the non-linearly realized SUSY in the VA model: it leads to a supergravity theory with spontaneously broken local supersymmetry \cite{Deser:1977uq, Bergshoeff:2015tra}.  This is best understood in unitary gauge, where the Goldstone fermion disappears, and one is left with an unusual supergravity action consisting of a Rarita-Schwinger field with an associated \textit{Lagrangian} mass term\footnote{In curved background the Rarita-Schwinger field acquires an effective mass that depends on the cosmological constant \cite{Townsend:1977qa,Bergshoeff:2015tra}. In particular, it becomes massless whenever the (A)dS local supersymmetry invariance is restored \cite{Townsend:1977qa}. }, coupled to gravity with a cosmological constant whose effective value (and sign) depends on both $a$ and the Rarita-Schwinger mass (cfr.\ section 4 \cite{Bergshoeff:2015tra}). To see how this reduces to \eqref{0-form-gauged}, let us start by describing how the the global minimal (A)dS superalgebra  \cite{vanNieuwenhuizen:2004rh} reduces under the IW contraction. For this limit, the relevant part thereof is given by:
\be
\ba \label{eq:superAdSalgebra}
    \{Q_\alpha, Q_\beta\} &= (\gamma^{m})_{\alpha \beta} P_m + \frac{1}{L} (\gamma^{m n})_{\alpha \beta} M_{m n}\comma \\
     [M_{m n}, Q_\alpha] &= \frac{1}{2} (\gamma_{m n})_\alpha{}^{\beta} Q_\beta\comma \; \;
[Q_\alpha, P_m] =  \frac{1}{L} (\gamma_m)_\alpha{}^{\beta} Q_\beta \comma\\
\ea
\ee
 where $L$ is just the AdS radius (the dS case just amounts to take $L^2\to -L^2$), and lower case letters indicate flat indices. 

Under the same IW contraction considered above $Q_\alpha = \omega S_\alpha$, the first two commutation relations in \eqref{eq:superAdSalgebra} reduce to \eqref{eq:0formalgebra}, while the latter gives: $[S_\alpha, P_m]= \frac{1}{L} (\gamma_m)_\alpha{}^{\beta} S_\beta$. This non-trivial commutation relation in the algebra, is the one responsible for the unusual gauge transformation for the Rarita-Schwinger background field we have proposed in \eqref{eq:modifiedpsimuvar}. 

This can be also checked at the level of field action. Consider the gravitino
$\psi_\mu$ gauge transformation dictated by  the gauged (A)dS supersymmetry algebra: \cite{vanNieuwenhuizen:2004rh}
\be \label{eq:AdSgravitinovariation}
\delta \psi_\mu = \partial_\mu \epsilon - \frac{1}{2L} e_\mu{}^n \gamma_n \epsilon + \frac{1}{4} \omega_\mu{}^{m n} \gamma_{m n} \epsilon \period
\ee
Under IW contraction,  we must rescale the gravitino field and the gauge parameter as $\psi_\mu \to \psi_\mu/\omega\comma \epsilon \to \epsilon/\omega$. Furthermore, being interested in confronting this with the gauged action on flat-space background \eqref{0-form-gauged},  we  take a weak-field limit of the gravitational field where $e_\mu{}^n \to \delta_\mu{}^\nu$ ($n \to \nu$) and $\omega_\mu{}^{m n} \to 0$. After this contraction and limit\footnote{See also a later article \cite{Zumino:1977av} that generalizes the original VA model on Minkowski spacetime to AdS spacetime and there the same mass term is needed even when the AdS global supersymmetry is ungauged.} \eqref{eq:AdSgravitinovariation} reproduces exactly the modified gauge variation we propose \eqref{eq:modifiedpsimuvar} upon identifying  the gauge coupling with the (A)dS radius\footnote{Note that this gives a precise relation between the parameter $f$ and $m$ in eq. (4.10) of \cite{Bergshoeff:2015tra}, $m^2 = \tfrac{2}{\sqrt{3}} f \kappa$ dictating a precise value of the effective cosmological constant.} as: 
\be
g = - \frac{d}{2L}\comma \qquad  a = - \frac{2L}{\sqrt{(d-1)(d-2)}}\period
\ee
While the weak-field limit is needed to confront directly with the flat-space system \eqref{0-form-gauged}, if one were to consider the gauging of the shifting symmetry on arbitrary curved manifolds, this would not be needed. Indeed, the transformation \eqref{eq:AdSgravitinovariation} corresponds to the modified gauge variation of the background gauge field on arbitrary backgrounds \eqref{curvedspacegauge} as described in Appendix B.

\paragraph{\textbf{Magnetic symmetries.}}
The free tensor-spinor fields described by the action \eqref{tensorspinor}, in addition to the symmetries generated by the (higher-form) currents \eqref{currents}, also possess
conserved $(p+1)$-form currents:
\be 
\dd{\bar{H}_{(p+1)}}= \dd{H_{(p+1)}} = 0 \comma\qquad H_{(p+1)} = \dd{\psi}_{(p)}\comma
\ee
as a consequence of Bianchi identity. They act as infinitesimal generators for a continuous \textit{magnetic} $(d-p-2)$-form symmetry, whose topological operators are labeled by a spinorial charge $\theta$ and topological choice of cycle $\mathcal{M}^{(p+1)}$:
\be
U^{(m)}_\theta(\mathcal{M}^{(p+1)}) = \exp\left( i \int_{\mathcal{M}^{(p+1)}} \left( \bar\theta H_{(p+1)} + \bar{H}_{(p+1)} \theta \right) \right)
\ee
The charged objects under this symmetry are disorder operators with prescribed singularity along $(d-p-2)$-cycles, inducing non-trivial holonomy along the dual intersecting cycles. This generalizes to this context the notion of \textit{'t Hooft operators}, well established in the bosonic case \cite{Gaiotto:2014kfa}.  As a prototypical example, take $d=4$, $p=1$ and consider  the line operator $T_m(\Gamma^{(1)})$ imposing the following prescribed singularity of the Rarita-Schwinger field: 
$\lambda_{(1)} = m (1-\cos\theta) \dd{\psi} + {\rm c.c.\ }\comma$
along the 1-dimensional cycle $\Gamma^{(1)}$. This line operator is readily shown to be charged under $U^{(m)}_\theta(S^{(2)})$ through the Ward Identity: 
\be 
U_\theta(S^{(2)})T_m(\Gamma^{(1)}) = e^{i(\bar \theta m + \bar{m} \theta)\langle \Gamma^{(1)}, S^{(2)}\rangle }   T_m(\Gamma^{(1)})\comma
\ee
valid inside any correlation function.
\paragraph{\textbf{Dual magnetic description.}}
In the bosonic case, it is often possible to find an equivalent descriptions of the theory, where disorder operators have a representation as holonomies of dual variables. For instance, in generalized Maxwell theory, this is achieved upon going to the dual EM frame, obtained by dualizing the $p$-form connection $A_{(p)} \leftrightarrow \hat{A}_{(d-p-2)}$,  where $F^{(p+1)}\to (\star F )^{(d-p-1)}$. Indeed, in this dual frame,  't Hooft operators are simply worldlines of the magnetic connection $\hat{A}_{(d-p-2)}$. Our proposal of magnetic symmetries in fermionic free theories motivates us to revisit the long-standing problem of the dualization procedure for fermionic fields. Here we exhibit a rather complete discussion, focusing on a Dirac spinor in four spacetime dimensions. The generalization to higher dimensions and form degree will be discussed in~\cite{Victor}.

The equivalence between a Dirac field $\psi$ and a fermionic 2-form $\chi_{(2)}$ is shown~\cite{Townsend:1979yv,Deser:1980fy} upon considering, as an intermediate step, the \textit{parent Lagrangian}:
\be \label{eq:parentLag}
\begin{split}
\mathcal{L}_{\text{parent}}= & \varepsilon^{\mu\nu\alpha\beta} \bar{\chi}_{\mu\nu}   \left( \partial_\alpha \psi_\beta + \gamma_\alpha \phi_\beta \right)   \\
& + \bar{\xi}^{\mu\nu} \partial_\mu \phi_\nu  + \bar{\psi}_\mu \gamma^{\mu\nu} \phi_\nu + \text{c.c.} \comma
\end{split}\ee
and introducing an extra gamma-traceless fermionic 2-form $\xi_{(2)}$. On one hand, by integrating out the two fields $\chi_{(2)}$ and $\xi_{(2)}$, one enforces constraints that are solved by: $\psi_\mu = \partial_\mu \alpha + \gamma_\mu \psi$, $\phi_\mu = \partial_\mu \psi$ for two Dirac spinors $\alpha$ and $\psi$; once those are substituted back into  \eqref{eq:parentLag}, one obtains the \textit{electric frame} free Dirac Lagrangian for the single spinor $\psi$. On the other hand, integrating out the fields $\psi_\mu$ and $\phi_\mu$ gives the equivalent \textit{magnetic frame} description of the free Dirac Lagrangian in terms of the degrees of freedom of the \textit{magnetic dual} $\chi_{(2)}$:
\be \begin{split}\label{duallagra}
\mathcal{L}_{\rm \scriptstyle dual}= & - 6 \partial_{[\mu} \bar{\chi}_{\alpha\beta]} \gamma^\mu \chi^{\alpha\beta} - \frac{2}{3} \partial_{\mu} \bar{\chi}_{\alpha\beta} \gamma^{\mu\alpha\beta} \gamma^{\rho \sigma} \chi_{\rho \sigma}  \\
&- \bar{\chi}_{\alpha\beta} \varepsilon^{\mu\nu\alpha\beta} \partial_\mu \partial^\lambda \xi_{\lambda \nu} + \text{c.c.} \comma
\end{split}\ee
where the last term should be regarded as a \textit{partial gauge fixing term} enforcing the correct amount of degrees of freedom on $\chi_{(2)}$. The latter enjoys the gauge freedom: $\delta \chi_{\alpha\beta} = 2 \partial_{[\alpha} \gamma_{\beta]} \epsilon$. We shall regard \eqref{duallagra} as a \textit{magnetic frame} description of a Dirac fermion. The unconventional form of the Lagrangian \eqref{duallagra}, compared to \eqref{tensorspinor}, is a consequence of having considered a $2$-form in $d=4$ (that does not satisfy $d \geq 2p+1$).

A direct relation between the Dirac spinor $\psi$ in the electric frame and its dual magnetic $2$-form $\chi_{(2)}$ can be inferred from the identity:
\be 
\frac{1}{2} \varepsilon^{\alpha\beta \rho \sigma} \gamma_{\rho \sigma} \partial_{[\mu} \chi_{\alpha\beta]}+ \frac{2}{3} \varepsilon_{\mu\nu\alpha\beta} \partial^\nu \chi^{\alpha\beta} = \phi_\mu 
 = \partial_\mu \psi\comma
\ee
that should be regarded as a fermionic analogous of its bosonic counterpart $\dd{\varphi} = \star \dd{B}_{(2)}$. 
In the magnetic frame one has the two conserved currents:
\be \begin{split} 
\mathcal{J}^{(e)}_\mu &= \varepsilon_{\mu\nu\alpha\beta} \partial^\nu \chi^{\alpha\beta}\\
\mathcal{J}^{(m)}_{\mu\nu\rho} &= 6 \gamma_{[\mu} \chi_{\nu\rho]} + \frac{2}{3} \gamma_{\mu\nu\rho}\gamma_{\sigma\tau} \chi^{\sigma\tau} - \varepsilon_{\mu\nu\rho\sigma} \partial_\lambda \xi^{\lambda \sigma}
\end{split}
\ee
that generate respectively a $0$-form and a $2$-form shifting symmetry: 
$\delta\chi_{\mu\nu} =  2 \gamma_{[\mu} \epsilon_{\nu]}$. 
The charged objects under the $2$-form symmetry, that were  't Hooft disorder operators in the \textit{electric frame}, in the magnetic frame admit a representation in terms of holonomies of $\chi_{(2)}$:
\be 
T_\theta(\mathcal{M}^{(2)}) = \exp(i\int_{\mathcal{M}^{(2)}}\left(\bar\theta \chi_{(2)} + \bar{\chi}_{(2)} \theta \right))\comma
\ee
charged under the action of the topological operators \be 
U_{\lambda}(\mathcal{S}^{(1)}) = \exp(i\int_{\mathcal{S}^{(1)}} \left(\bar\lambda \star\mathcal{J}^{(m)}_{(3)} + \star\bar{\mathcal{J}}^{(m)}_{(3)} \lambda\right))\comma\ee
as easily verifiable using the Ward Identity associated to $\mathcal{J}_{(3)}^{(m)}$. The opposite holds true for the $0$-form symmetry whose charged objects in the magnetic frame are disorder operators. 

\paragraph{\textbf{Fermionic Higher-group structures.}}
In recent years, there has been a consistent effort devoted to generalize the algebraic structure behind bosonic symmetry beyond ordinary $1$-group case, cfr.\ e.g.\ \cite{Kapustin:2013uxa,Cordova:2018cvg,Iqbal:2020lrt,Bartsch:2023pzl} and references therein.
Here, we explore higher-algebraic structures of fermionic symmetries. We present an example of $2$-group fermionic (gauge) symmetry in a free massive $2$-form spinor $\Psi_{\mu \nu}$ in fixed (A)dS$_d \ (d\ge 5)$ background coupled to an additional Goldstone field $\Phi_{\mu}$. This system,  also previously considered in \cite{Zinoviev:2009wh}, can be described by the Lagrangian:
\begin{widetext}
\be\label{eq:order2fermionAntisymLag}
\mathcal{L}= - \bar{\Psi}_{\mu \nu} \gamma^{\mu \nu \alpha \beta \gamma} \nabla_\alpha \Psi_{\beta \gamma}+ m \bar{\Psi}_{\mu \nu} \gamma^{\mu \nu\alpha \beta} \Psi_{\alpha \beta}-\Phi_\mu \gamma^{\mu \nu \rho} \nabla_\nu \Phi_\rho
 +b_2 \bar{\Phi}_\mu \gamma^{\mu \nu} \Phi_\nu+b_1\left(\bar{\Psi}_{\mu \nu} \gamma^{\mu \nu \alpha} \Phi_\alpha+ \bar{\Phi}_\alpha  \gamma^{\alpha \mu \nu } \Psi_{\mu \nu} \right) \comma
\ee
\end{widetext}
where $\nabla_\mu$ is the (A)dS covariant derivative.  This model admits a fermionic shifting $2$-group gauge symmetry
\be\begin{split}\label{eq:fermion2-gauge}
\delta \Phi_\mu &= \nabla_\mu \eta -2b_1 \xi_\mu + \frac{b_2}{d-2} \gamma_\mu \eta \\
\delta \Psi_{\mu\nu} &=  2\nabla_{[\mu}\xi_{\nu]} -\frac{2m}{d-4} \gamma_{[\mu} \xi_{\nu]} - \frac{b_1}{(d-3)(d-4)}\gamma_{\mu\nu}\eta \comma
\end{split}\ee
whenever the parameters $b_1,b_2$, the mass $m$, and radius of (A)dS $L$ ($L^2\to - L^2$ for dS), are related to each other by:
\be
b_2  =-\frac{d-2}{d-4}m\comma \quad m^2 = \frac{d-4}{d-3}b_1^2 -\frac{(d-4)^2 }{8L^2} \period
\ee
Exactly as in in the bosonic case (cfr.\ \cite{baez2004higher}), the structure of fermionic 2-group gauge transformation allows us to introduce a gauge-invariant \textit{fake-curvature}, taking the form
\be
\mathcal{F}_{\mu\nu} = \nabla_{[\mu} \Phi_{\nu]} + b_1 \Psi_{\mu\nu} - \frac{m}{d-4} \gamma_{[\mu}\Phi_{\nu]} \ ,
\ee
which can serve as a building block for 2-gauge-invariant actions. 
This structure is not limited to this specific example, but is rather general: it is straightforward to generalize this discussion \cite{Zinoviev:2009wh} to coupled system of $p$ and $(p-1)$-form spinors on  (A)dS$_d$ backgrounds with  $(d\ge 2p+1)$. 

\paragraph{\textbf{Outlook.}}

In this letter we explored various aspects of fermionic symmetries of free tensor-spinor fields, including free Dirac fermions. 
Many interesting points remain open to investigation.
First of all, it would be worth studying examples of interacting theories exhibiting fermionic higher-form symmetries. Following the discussions of the VA model, it appears natural to investigate whether non-trivial fermionic generalized symmetries survive in the full supergravity theory with non-linear realization of SUSY (e.g.\ in \cite{Bergshoeff:2015tra}), that is the (A)dS supergravity uplift of the VA model reduction. Moreover, with regards to possible physical models realizing these symmetries, the novel St\"{u}ckelberg mechanism we propose to give mass to  spinors might find non-trivial applications in particle physics or condensed matter physics that are worth exploring. A further natural setting to study fermionic symmetries would be the supersymmetric version of Vasiliev higher-spin-gravity, and their realization in CFT dual theory \cite{Vasiliev:2004cm,Giombi:2012ms}. 
In this letter we started exploring the realm of higher-categorical structure by considering some examples of fermionic higher-groups, and the landscape of possible algebraic structures is yet to be fully unveiled. For instance, by now many instances  of bosonic non-invertible symmetries are known (\cite{Shao:2023gho, Schafer-Nameki:2023jdn} and references therein) in arbitrary dimensions. It is intriguing to study whether similar constructions admit a fermionic generalization.

\paragraph{\textbf{Acknowledgments.}}
We thank Clay Cordova, Stefano Cremonesi, Qiang Jia, Ho Tat Lam, Craig Lawrie, Victor Lekeu, Teng Ma, Marwan Najjar, Jun Nian, Yi Pang, Ziqi Yan, Jinwu Ye for discussions. RL, YNW and YZ are supported by National Natural Science Foundation of China under Grant No. 12175004, by Peking University under startup Grant No. 7100603667, and by Young Elite Scientists Sponsorship Program by CAST (2022QNRC001, 2023QNRC001).
YZ is supported by National Science Foundation of China under Grant No. 12305077 and also by the Office of China Postdoc Council (OCPC) and Peking University under Grant No. YJ20220018.

\bibliography{biblio}

\onecolumngrid

\appendix

\section{Appendix A: Conventions}
\label{sec:conventions}
We inherit most of the conventions from \cite{Wang:2023iqt}. We use the “mostly plus” signature for $d$-dimensional Minkowski metric: $\eta = \text{diag} (-,+,\ldots,+) $.
Gamma matrices $\gamma_\mu$ ($\mu=0,\ldots,d-1$) satisfy the anti-commutation relation
\be
\{\gamma_\mu, \gamma_\nu \} = 2 \eta_{\mu\nu} \, .
\ee
Hermitian property of gammas is $(\gamma^\mu)^\dagger = \gamma^0 \gamma^\mu \gamma^0$.
In $d = 2 m$ dimensions the chirality matrix is defined as 
\be \label{eq:chiralityelement}
\gamma_{d+1} = (-i)^{m+1} \gamma_0 \gamma_1 \ldots \gamma_{d-1} \,.
\ee 
The antisymmetric summations of $\gamma$-matrices are defined as
\be  \gamma^{\mu_1\mu_2\dots \mu_r} = \gamma^{[\mu_1}\gamma^{\mu_2}\cdots \gamma^{\mu_r]} \equiv \frac{1}{r!}\sum_{\kappa\in S_r} {\rm sgn}(\kappa) \  \gamma^{\mu_{\kappa(1)}}\gamma^{\mu_{\kappa(2)}}\cdots \gamma^{\mu_{\kappa(r)}} \, . \ee 
In components, we have the gamma matrices $p$-forms 
\be 
\gamma_{(p)}= \frac{1}{p!} \gamma_{\mu_1 \ldots \mu_p} \dd x^{\mu_1} \wedge \ldots \wedge \dd x^{\mu_p} \ .
\ee
One can deduce the useful identity:
\be
\gamma^{\mu_1 \dots \mu_r \nu_1\dots\nu_k}\gamma_{\nu_k\dots \nu_1} = \frac{(d-r)!}{(d-r-k)!} \gamma^{\mu_1\dots \mu_r} \,.
\ee
For a spinor $\psi$, its Dirac conjugate is $\bar{\psi} = i \psi^\dagger \gamma^0$. \\
A differential $p$-form $\omega_{(p)}$ is expressed in components as 
\be
\omega_{(p)} = \frac{1}{p!} \omega_{\mu_1 \ldots \mu_p} dx^{\mu_1} \wedge \ldots \wedge dx^{\mu_p} \, ,
\ee
and its exterior derivative $(d \omega)_{(p+1)}$ is a $(p+1)$-form with components
\be
 (d \omega )_{\mu_1\ldots\mu_{p+1}} = (p+1) \ptl_{[\mu_1} \omega_{\mu_2\ldots\mu_{p+1]}}  \, .  
\ee
The components of the wedge product of a $p$-form $\omega_{(p)}$ and a $q$-form $\eta_{(q)}$ are
\be
 (\omega \wedge \eta)_{\mu_1\ldots\mu_p\nu_1\ldots\nu_q} = \frac{(p+q)!}{p!q!} \omega_{[\mu_1\ldots\mu_p} \eta_{\nu_1\ldots\nu_q]}\, ,
\ee
and similarly for $p$-form spinors,
\be
\psi_{(p)} = \frac{1}{p!} \psi_{\mu_1 \ldots \mu_p} \dd x^{\mu_1} \wedge \ldots \wedge \dd x^{\mu_p} \ .
\ee
The Hodge star operator $\star$ maps $p$-forms to $(d-p)$-forms and our convention is  
\be
(\star \omega)_{\mu_1\ldots\mu_{d-p}} = \frac{1}{p!} \varepsilon_{\mu_1\ldots\mu_{d-p}}{}^{\nu_1\ldots\nu_p} \omega_{\nu_1\ldots\nu_p} \, ,
\ee where $\varepsilon_{\mu_1\ldots\mu_d}$ is the Levi-Civita symbol and $\varepsilon_{0 1\ldots d-1} = 1$, $\varepsilon^{0 1\ldots d-1} = -1$.\\
For odd dimensions 
\be
   \gamma^{\mu_1\ldots\mu_p} = i^{\frac{d+1}{2}} \frac{1}{(d-p)!} \varepsilon^{\mu_1\ldots\mu_p\nu_1\ldots\nu_{d-p}} \gamma_{\nu_{d-p}\ldots\nu_1} \,,
\ee
while for even $d$, this identity becomes
\be
   \gamma^{\mu_1\ldots\mu_p} \gamma_{d+1} = - (-i)^{\frac{d}{2}+1} \frac{1}{(d-p)!} \varepsilon^{\mu_p\ldots\mu_1\nu_1\ldots\nu_{d-p}} \gamma_{\nu_1\ldots\nu_{d-p}} \,.
\ee\\
On a curved manifold with metric $g_{\mu\nu}$, the Levi-Civita symbol $\varepsilon$ generalizes to a tensor according to the normalization
\be
\varepsilon_{0 1\ldots d-1} = \sqrt{|\det{g}|} \,,\qquad \varepsilon^{0 1\ldots d-1} = \frac{-1}{\sqrt{|\det{g}|}} \,. 
\ee
The invariant volume form is 
\be
\sqrt{|\det{g}|} \, d^dx \equiv \sqrt{|\det{g}|}\, dx^0 \wedge \ldots \wedge dx^{d-1} = \frac{1}{d!}\varepsilon_{\mu_1\ldots\mu_d} \, dx^{\mu_1} \wedge \ldots \wedge dx^{\mu_d} \, , 
\ee
here we would also write $dV^{\mu_1\ldots\mu_d} \equiv dx^{\mu_1} \wedge \ldots \wedge dx^{\mu_d}$ for short. \\
Integration of $d$-forms over the manifold $\mathcal{M}^{(d)}$ is given as 
\be
\ba
\int_{\mathcal{M}^{(d)}}  \upsilon_{(d)} &= \int_{\mathcal{M}^{(d)}}  \frac{1}{d!} \upsilon_{\mu_1\ldots\mu_d} \, dx^{\mu_1} \wedge \ldots \wedge dx^{\mu_d} = \int_{\mathcal{M}^{(d)}}  \frac{1}{d!} \upsilon_{\mu_1\ldots\mu_d} \, dV^{\mu_1\ldots\mu_d} \cr  
&= \int_{\mathcal{M}^{(d)}} \upsilon_{01\ldots d-1} \, d^dx \equiv \int_{\mathcal{M}^{(d)}} \upsilon(x)_{01\ldots d-1} dx^0 dx^1 \ldots dx^{d-1} \, ,\cr 
\ea
\ee
and integration of a scalar ($0$-form) $\phi$ is defined as the integral of its Hodge dual 
\be
\int_{\mathcal{M}^{(d)}} \star \phi = \int_{\mathcal{M}^{(d)}} \phi \, \sqrt{|\det{g}|} \, d^dx \, . 
\ee
Useful formulae: 
\be
\ba
\star \omega \wedge \eta &= \star \eta \wedge \omega= \frac{1}{p!} \omega_{\mu_1\ldots\mu_p} \eta^{\mu_1\ldots\mu_p} \sqrt{|\det{g}|} \, d^dx \, ,\cr
d \star \upsilon \wedge \omega &= (-1)^{d-p-1} \frac{1}{p!} \ptl_\mu \upsilon^{\mu \nu_1 \ldots \nu_p} \omega_{\nu_1\ldots\nu_p} \sqrt{|\det{g}|} \, d^dx \, , \cr 
\ea
\ee
where $\omega$ and $\eta$ are both $p$-forms and $\upsilon$ is a $(p+1)$-form.

For submanifolds $U$ and $V$ with dimension $p$ and $d-p-1$ and such that $V$ is the boundary of a $(d-p)$-dimensional submanifold $W$, i.e. $\ptl W= V$ (in fact, both $U$ and $V$ should be boundaries of some other submanifolds in order to define the linking number). 
The linking number $\langle U, V\rangle$ is given as the intersection number $\mc{I}(U,W)$ of $U$ and $W$
\be \label{eq:deflinking}
 \langle U, V\rangle = \mc{I}(U,W) = \int_{\mathcal{M}^{(d)}} J_{(d-p)}(U) \wedge J_{(p)}(W) = \int_{U} J_{(p)}(W) \ .
\ee

\section{Appendix B: Fermionic Symmetries on Curved Manifolds}
\label{sec:curved}

We discuss the presence of fermionic higher-form symmetries on a fixed curved manifold $\mc{M}^{(d)}$, that we must endow with a spin structure. We first consider the fermionic shifting symmetry of a free Dirac spinor $\psi$, which has the action
\be
S[\psi]=-\int_{\mathcal{M}^{(d)}}e\bar{\psi}\gamma^\mu\nabla_\mu\psi\,.
\ee
The theory still have a fermionic symmetry $\delta\psi=\epsilon$, where the spinor parameter $\epsilon$ obeys 
\be
\label{cov-cons-spinor}
\nabla_\mu\epsilon=0\period
\ee
This entails that a shifting symmetry only exists on manifolds $\mc{M}^{(d)}$ with covariantly constant spinors. We can also derive this result from the operator $U_\epsilon(\mc{M}^{(d-1)})$, associated with this shifting symmetry:
\be
U_\epsilon(\mathcal{M}^{(d-1)})=\exp(i\int_{\mathcal{M}^{(d-1)}}\bar{\epsilon}\star\mc{J}_{(1)}+\text{c.c.}\ )\comma
\ee
that is only topological only if 
\be
\ba
\dd{(\bar{\epsilon}\star\mc{J}_{(1)}+\text{c.c.})}&=\cr
\nabla(\bar{\epsilon}\wedge\star\mc{J}_{(1)}+\text{c.c.})&=\cr
\nabla\bar{\epsilon}\wedge\star\mc{J}_{(1)}+\bar{\epsilon}\wedge\nabla(\star\mc{J}_{(1)})+\text{c.c.}\ &=0 \comma
\ea
\ee
which is only possible when $\nabla_\mu\epsilon=0$ and $\nabla_\mu\bar\epsilon=0$.

For different spacetime background $\mc{M}_d$, the dimension of solution space to (\ref{cov-cons-spinor}) is also different. This is a unique feature of fermionic symmetry which is different from the bosonic case, where the global symmetries defined by topological currents are background independent. 

The argument can be applied to fermionic $p$-form symmetries as well. Consider the generator 
\be
U_\epsilon(\mathcal{M}^{(d-p-1)})=\exp\left(i\int_{\mathcal{M}^{(d-p-1)}}\bar{\epsilon}\wedge\star\mc{J}_{(p+1)}+\text{c.c.}\right)\,.
\ee
It is only topological if
\be
\dd{(\bar{\epsilon}\wedge\star\mc{J}_{(p+1)}+\text{c.c.})}=
\nabla\bar{\epsilon}\wedge\star\mc{J}_{(p+1)}+\bar{\epsilon}\wedge\nabla(\star\mc{J}_{(p+1)})+\text{c.c.}=0\period
\ee

We can also study systems of fermions compact spin manifolds $\mc{M}^{(d)}$ that admit more than one spin structure. For example, we can consider a free Dirac spinor $\psi$ on a torus $T^d$; this manifold admits $2^d$ spin structures labeling the periodic/anti-periodic boundary conditions for the fermionic degrees of freedom around each cycle $S_i^1\subset T^d$. Also in the case of anti-periodic boundary conditions around a circle $S_i^1$: $\psi(x+2\pi)=-\psi(x)$, the fermionic shifting symmetry $\psi\rightarrow\psi+\epsilon$ remain unbroken as, for consistency,  we must impose the same boundary conditions on the symmetry parameter $\epsilon(x+2\pi)=-\epsilon(x)$. This holds more generally for fermionic $p$-form symmetries. Indeed,  since all topological operators: ($\chi_{(p)}$ represents some fermionic $p$-form here)
\be
U_\epsilon(\mc{C}^{(p)})=\exp(i\int_{\mc{C}^{(p)}}\bar\epsilon \chi_{(p)}+\text{c.c.})\comma
\ee
and the all charged operators:
\be
V_\eta(\mc{C}^{(d-p-1)})=\exp(i\int_{\mc{C}^{(d-p-1)}}\bar\eta \psi_{(d-p-1)}+\text{c.c.})
\ee
are built out of fermionic bilinears, they are still well-defined under anti-periodic boundary conditions.

Finally, we briefly comment on the gauging of fermionic shifting symmetries on a curved spacetime with covariantly constant spinors. As in flat space, we promote the global shifting symmetry of a $p$-form tensor spinor to a local one, by introducing a $(p+1)$-form gauge field with modified gauge transformation:
\be\label{curvedspacegauge}
\delta \lambda_{\mu_1\cdots\mu_{p+1}} = \nabla_{[\mu_1}{\epsilon_{\mu_2\cdots \mu_{p+1}]}} + \frac{g}{d} \gamma_{[\mu_1} \epsilon_{\mu_2,\cdots \mu_{p+1}]}\comma \quad \delta \psi = \alpha g \, \epsilon \comma
\ee
where $\nabla_\mu$ is the covariant derivative on (A)dS. 
A direct generalization of what presented in the main text allow us to deduce that the minimally coupled action compatible with the gauge transformations \eqref{curvedspacegauge}, can be simply obtained by promoting all the partial derivatives to covariantized ones in \eqref{p-form-gauged}. 
Just for the sake of clarity, we report here the final form of the Lagrangian in components:
\be \begin{split}\label{gaugedcationcurved}
\mathcal{L} =  &- \bar{\lambda}_{\mu_1\cdots\mu_p} \gamma^{\mu_1\cdots \mu_p \nu \rho_1\cdots\rho_p} \nabla_\nu 
 \lambda_{\rho_1\cdots\rho_p} - 
 (-1)^p g \frac{d-2p}{d} \bar{\lambda}_{\mu_1\cdots\mu_p} \gamma^{\mu_1\cdots\mu_p \rho_1\cdots\rho_p}\lambda_{\rho_1\cdots\rho_p} 
  \\ 
 &-\bar{\psi}_{\mu_1\cdots\mu_{p+1}} 
 \gamma^{\mu_1\cdots \mu_{p+1} \nu \rho_1\cdots\rho_{p+1}} \nabla_\nu  \lambda_{\rho_1\cdots\rho_p}  -(-1)^{p+1} g \frac{d-2p-2}{d} \bar{\psi}_{\mu_1\cdots\mu_{p
 +1}} \gamma^{\mu_1\cdots\mu_{p+1} \rho_1\cdots\rho_{p+1}}\psi_{\rho_1\cdots\rho_{p+1}} \\
 &+ \bar{\psi}_{\mu_1 \cdots \mu_{p+1}} {\mathcal{J}}^{\mu_1 \cdots \mu_{p+1}}  + \overline{\mathcal{J}}^{\mu_1 \cdots \mu_{p+1}} \psi_{\mu_1 \cdots \mu_{p+1}}  \, ,
\end{split}
\ee
On arbitrary curved backgrounds, this action is invariant for any values of two scales: the radius of (A)dS $L$ ($L^2\to -L^2$ for  dS) and the gauge coupling strength $g$. This matches exactly the linearized version of the (A)dS supergravity with spontaneously broken SUSY presented in \cite{Bergshoeff:2015tra} (cfr.\ with section 4 thereof).The one we present in the main text has only a single scale $g$, as we decided to gauge the shifthing symmetry in flat space. This fixes the effective cosmological constant to be zero (see footnote 8 in the main text).

\section{Appendix C: Hamiltonian analysis of coupled systems}
We present a detailed calculation and analysis of the degree of freedom in both the massive Rarita-Schwinger theory and the theory minimally coupling Rarita-Schwinger field with a spinor field as \eqref{0-form-gauged}. The degrees of freedom of the two theories are indeed equal to each other, corroborating our assertion that the two theories are equivalent.

Written in a form such that the time direction $x^0$ and the spatial directions $x^i$ $(i=1,\dots,d-1)$ are separated, theory \eqref{0-form-gauged} reads
\be\ba
\label{0-form-gauged-explicit}
S_{\rm gauged} =\int d^dx \ \mathcal{L} = \int d^d x \ \Big[   &-\bar{\psi}(-\gamma^0\ptl_0 + \gamma^i\ptl_i)\psi - g\bar{\psi}\psi + \frac{d-2}{d}g(-\bar{\lambda}_0 \gamma^{0i} \lambda_i + \bar{\lambda}_i\gamma^{i0}\lambda_0) \\
&+\big( \bar{\lambda}_0\gamma^{0ij}\ptl_i\lambda_j + \bar{\lambda}_i\gamma^{i0j}\ptl_0\lambda_j +  
\bar{\lambda}_i\gamma^{ij0}\ptl_j\lambda_0 -
\bar{\lambda}_i\gamma^{ijk}\ptl_j\lambda_k \big) \\
&+ \alpha g \bar{\psi}(-\gamma^0\lambda_0+ \gamma^i\lambda_i) -\alpha g (-\bar{\lambda}_0\gamma^0\psi + \bar{\lambda}_i\gamma^i\psi)  \Big]  \ ,
\ea\ee
with off-shell gauge invariance under \eqref{eq:modifiedpsimuvar}.
The Euler-Lagrange equations gives
\be
\ba
\label{EL}
&\frac{\delta \mathcal{L}}{\delta\bar{\lambda}_0} - \ptl_\mu\frac{\delta \mathcal{L}}{\delta\big(\ptl_\mu \bar{\lambda}_0\big)}=-\frac{d-2}{d}g\gamma^{0i} \lambda_i + \gamma^{0ij}\ptl_i\lambda_j + \alpha g\gamma^0\psi = 0 \ , \\
&\frac{\delta \mathcal{L}}{\delta\bar{\lambda}_i} - \ptl_\mu\frac{\delta \mathcal{L}}{\delta\big(\ptl_\mu\bar{\lambda}_i\big)}=\frac{d-2}{d}g\gamma^{i0}\lambda_0 + \gamma^{i0j}\ptl_0\lambda_j +  \gamma^{ij0}\ptl_j\lambda_0 -\gamma^{ijk}\ptl_j\lambda_k-\alpha g\gamma^i\psi = 0 \ , \\
&\frac{\delta \mathcal{L}}{\delta\bar{\psi}} - \ptl_\mu\frac{\delta \mathcal{L}}{\delta\big(\ptl_\mu\bar{\psi}\big)}= (\gamma^0\ptl_0 - \gamma^i\ptl_i)\psi -g\psi + \alpha g(-\gamma^0\lambda_0 + \gamma^i\lambda_i) = 0 \ . 
\ea
\ee
The variation of $\bar{\lambda}_0$ generates the constraint
\be  \mathcal{K} \equiv  -\frac{d-2}{d}g\gamma^{0i} \lambda_i + \gamma^{0ij}\ptl_i\lambda_j + \alpha g\gamma^0\psi \ , \ee
from which we can write the Lagrangian into a sum of constraint, time derivative and Hamiltonian,
\be\ba  
\mathcal{L} = \ &\bar{\lambda}_0\mathcal{K} +   \mathcal{D} - \mathcal{H}  \ , \\
{\rm time \ derivative:\ }\mathcal{D} = \  &\bar{\psi}\gamma^0\ptl_0\psi + \bar{\lambda}_i \gamma^{i0j}\ptl_0\lambda_j \ . 
\ea\ee
The canonical momenta are
\be
P_{\lambda_i} = \bar{\lambda}_k\gamma^{k0i} =i \lambda_k^\dagger \gamma^{ki} \ , \ P_{\psi} = \bar{\psi}\gamma^0  =  -i\psi^\dagger \ ,
\ee
yielding the equal time commutation relations:
\be\ba
\{  \lambda_{i\alpha} (x^0,\Vec{x}), P_{\lambda_j \beta} (y^0,\Vec{y}) \}  &= i\delta_{\alpha\beta} \delta_{ij}\delta(x^0-y^0)\delta^{(3)}(\Vec{x}-\Vec{y}) \ , \\
\{  \psi_\alpha (x^0,\Vec{x}) , P_{\psi\beta} (y^0,\Vec{y})\}&= i\delta_{\alpha\beta} \delta(x^0-y^0)\delta^{(3)}(\Vec{x}-\Vec{y}) \ .
\ea\ee
The above anti-commutation relations give rise to
\be \{  \mathcal{K},\mathcal{H} \} = 0 \ , \ee
indicating that $\mathcal{K}$ is a first class constraint. The formula for calculating the classical degrees of freedom is \cite{9d50b89c-daf8-39a3-a319-7c19d071d0ae,Adler_2017}
\be  \frac{1}{2}\big( N - 2F-S  \big)  \ee
where $N$ is the number of real canonical variables, $F$ is the number of real first class constraint and $S$ is the number of real second class constraint. In our case, $N = 2\left((d-1)2^{[d/2]} + 2^{[d/2]} \right)$, $ F = 2^{[d/2]}$ (one for each spinor component of the constraint $\mathcal K$), and  $S=0$. Hence:
\be \label{dofcoupled}
{\rm DOF}= \frac{1}{2}\left(d \cdot 2^{[d/2]} -2 \cdot 2^{[d/2]} \right) = \frac{1}{2} \left( d-2\right) 2^{[d/2]} \period
\ee

On the other hand, the massive Rarita-Schwinger theory is described by
\be\ba 
S &= -\int d^d x \ \bar{\lambda}_\mu \Big[ 
 \gamma^{\mu\nu\rho}\ptl_\nu - m \gamma^{\mu\rho} \Big]\lambda_\rho  \\
&= \int d^d x \ \Big[\bl_0 \gamma^{0ij}\ptl_i\lambda_j + \bl_i\gamma^{i0j}\ptl_0\lambda_j + \bl_i\gamma^{ij0}\ptl_j\lambda_0 -\bl_i\gamma^{ijk}\ptl_j\lambda_k + m\big( 
 -\bl_0\gamma^{0i}\lambda_i - \bl_i\gamma^{i0}\lambda_0 + \bl_i \gamma^{ij}\lambda_j \big)\Big] \ ,
\ea\ee
with Euler-Lagrange equation
\be\ba
&\frac{\delta \mathcal{L}}{\delta\bar{\lambda}_0} - \ptl_\mu\frac{\delta \mathcal{L}}{\delta\big(\ptl_\mu \bar{\lambda}_0\big)}= \gamma^{0ij}\ptl_i\lambda_j - m\gamma^{0i}\lambda_i = 0 \ , \\
&\frac{\delta \mathcal{L}}{\delta\bar{\lambda}_i} - \ptl_\mu\frac{\delta \mathcal{L}}{\delta\big(\ptl_\mu\bar{\lambda}_i\big)}=\gamma^{i0j}\ptl_0\lambda_j + \gamma^{ij0}\ptl_j\lambda_0 -\gamma^{ijk}\ptl_j\lambda_k -m \big( \gamma^{i0}\lambda_0 + \gamma^{ij}\lambda_j  \big) = 0 \ .
\ea\ee
The massive Rarita-Schwinger field in $d$ space-time dimensions has  degrees of freedom\cite{Freedman_VanProeyen_2012}:
\be  {\rm DOF} = \frac{1}{2}(d-2)2^{[d/2]}  \ . \ee
matching  the ones of the coupled system \eqref{dofcoupled}. 

\end{document}